\documentclass[aps,preprintnumbers,nofootinbib]{revtex4}
\usepackage{amsmath,amssymb}
\usepackage{graphicx}

\begin{document}
\preprint{MIT-CTP/3997}
\title[New Kinds of Quantum Statistics]{New Kinds of Quantum Statistics}

\author{Frank Wilczek}
\affiliation{Dept.~of Physics\\ Massachusetts Institute of Technology\\ Cambridge, MA 02139, USA}
\email{wilczek@mit.edu}

\begin{abstract}
I review the quantum kinematics of identical particles, which suggests new possibilities, beyond bosons and fermions, in 2+1 dimensions; and how simple flux-charge constructions embody the new possibilities, leading to both abelian and nonabelian anyons.   I briefly allude to experimental realizations, and also advertise a spinor construction of nonabelian statistics, that has a 3+1 dimensional extension. 
\end{abstract}

\maketitle

In quantum theory the notion of identity reaches a new level of precision and has profound dynamical significance.  It becomes important that two particles can be {\it precisely\/} identical, i.e. indistinguishable, as opposed to merely similar.   When passing from a classical description of indistinguishable particles to a quantum description one must supply additional rules, known as the quantum statistics of the particles.     

For many years it was thought that there are only two possibilities for quantum statistics: bosons and fermions.  But in 1977 Leinaas and Myrheim \cite{Leinaas} demonstrated, at the level of particle quantum mechanics, that there were additional theoretically consistent possibilities.   I'm told there were hints of this in earlier work in axiomatic field theory \cite{DR}.   The work of Leinaas and Myrheim received little attention, and their insight was rediscovered, independently, by Goldin, Menikoff, and Sharp \cite{GMS}, who realized it in the context of a special formulation of quantum mechanics using currents and densities;  and by me using conventional quantum field theory \cite{anyonsFW} (involving solitons, as below, and/or Chern-Simons terms \cite{chernSimonsFW}).   This circle of ideas came to life as physics in 1984, when Arovas, Schrieffer and I demonstrated \cite{qHallAnyons} -- theoretically, but I think quite convincingly -- that quasiparticles in the fractional quantum Hall effect obey forms of the new, ``anyon'' quantum statistics.  (That possibility was foreseen by Halperin \cite{qHallHalperin}.).   The anyonic behavior of quasiparticles (and quasiholes)  in the fractional quantum Hall effect is so closely integrated into the overall theory of those states that it can be subtle to demonstrate as an independent phenomenon.   A recent series of impressive experiments by V. Goldman and his collaborators \cite{goldman} have been interpreted this way, and other experiments, requiring less interpretation, are in the works.   

Rich mathematical possibilities arise when we consider {\it nonabelian\/} statistics.   In the abelian case the operations characteristic of quantum statistics -- roughly speaking: slow, distant exchange of particle positions -- are implemented as multiplications of the wave-function by a complex number (phase).  In the nonabelian case complex motions in large Hilbert spaces of degenerate states can come into play.   The possibility of exploiting a  robust mapping from operations in physical space (characterized topologically) to navigate through large Hilbert spaces has inspired visions of a possible route to quantum computing, known as topological quantum computing.   Physical realization of topological quantum computing is still far off, if it can be achieved at all, but the program has inspired impressive work, both theoretical and experimental.   An upcoming milestone may be demonstration of a proposal by Moore and Read \cite{mooreRead} that quasiparticles in an observed $\nu = \frac{5}{2}$ quantum Hall state obey nonabelian statistics.  Experimental programs to test this are well advanced, as well.  

Here I will describe a few of the most fundamental concepts underlying these developments in what might appear, to a quantum field theorist, as their simplest natural context.  (I will mention quantum Hall physics, experimental aspects, and quantum computing, but I will not even begin to do them justice.)    In the course of this review a few intriguing new ideas will come up, too.

\section{Braids, Permutations, and In Between}

Traditionally, the world has been divided between bosons (Bose-Einstein statistics) and fermions (Fermi-Dirac statistics).   Let's recall what these are, and why they appear to exhaust the possibilities.  

If two identical particles start at positions $(A, B)$ and transition to  $(A^\prime, B^\prime)$, we must consider both $(A, B) \rightarrow (A^\prime, B^\prime)$ and $(A, B) \rightarrow (B^\prime, A^\prime)$
as possible accounts of what has happened.  According the rules of quantum mechanics, we must add the amplitudes for these possibilities, with appropriate weights.   The rules for the weights encode the dynamics of the particular particles involved, and a large part of what we do in fundamental physics is to determine such rules and derive their consequences.   

In general, discovering the rules involves creative guesswork, guided by experiment.  One important guiding principle is correspondence with classical mechanics.  If we have a classical Lagrangian $L_{\rm cl.}$, we can use it, following Feynman, to construct a path integral, with each path weighted by a factor 
\begin{equation}\label{classicalWeight}
e^{i\int dt L_{\rm cl.}} \equiv e^{iS_{\rm cl.}}
\end{equation}
where $S_{\rm cl.}$ is the classical action.  This path integral provides -- modulo several technicalities and qualifications --  amplitudes that automatically implement the general rules of quantum mechanics.  Specifically: it sums over alternative histories, takes products of amplitudes for successive events, and generates unitary time evolution.   

The classical correspondence, however, does not instruct us regarding the relative weights for trajectories that are topologically distinct, i.e. that cannot be continuously deformed into one another.   Since only small variations in trajectories are involved in determining the classical equations of motion,  from the condition that $S_{\rm cl.}$ is stationary, the classical equations cannot tell us how to interpolate between topologically distinct trajectories.  We need additional, essentially quantum-mechanical rules for that.   

Now trajectories that transition $(A, B) \rightarrow (A^\prime, B^\prime)$ respectively $(A, B) \rightarrow (B^\prime, A^\prime)$ are obviously topologically distinct.   The traditional additional rule is: for bosons, add the amplitudes for these two classes of trajectories\footnote{As determined by the classical correspondence, or other knowledge of the interactions.}; for fermions, subtract.  

These might appear to be the only two possibilities, according to the following (not-quite-right) argument.  Let us focus on the case  $A = A^\prime, B = B^\prime$.   If we run an ``exchange'' trajectory $(A, B) \rightarrow (B, A)$ twice in succession, the doubled trajectory is a direct trajectory.   The the square of the factor we assign to the exchange trajectory must be the square of the (trivial) factor $1$ we associate to the direct trajectory, i.e. it must be $\pm 1$.  

This argument is not conclusive,  however, because there can be additional topological distinctions among trajectories, not visible in the mapping between endpoints. This distinction is especially important in 2 spatial dimensions, so let us start there.   (I should recall that quantum-mechanical systems at low energy can effectively embody reduced dimensionality, if their dynamics is constrained below an energy gap to exclude excited states whose wave functions have structure in the transverse direction.)   The topology of trajectory space is then specified by the {\it braid group}.   Suppose that we have $N$ identical particles.  Define the elementary operation $\sigma_j$ to be the act of taking particle $j$ over particle $j+1$, so that their final positions are interchanged, while leaving the other particles in place.  (See Figure 1.)  We define products of the elementary operations by performing them sequentially.  Then we have the obvious relation 
\begin{equation}\label{distantCommutation}
\sigma_j \sigma_k \, = \, \sigma_k \sigma_j; \, \, \, \, |j-k| \geq 2
\end{equation}
among operations that involve separate pairs of particles.   
We also have the less obvious {\it Yang-Baxter\/} relation
\begin{equation}\label{YangBaxter}
\sigma_j \sigma_{j+1} \sigma_j \, = \, \sigma_{j+1} \sigma_j \sigma_{j+1}
\end{equation}
which is illustrated in Figure 1.  The topologically distinct classes of trajectories are constructed by taking products of $\sigma_j$s and their inverses, subject only to these relations.   


\begin{figure}[ht]\label{figure1}
\includegraphics[width=8cm]{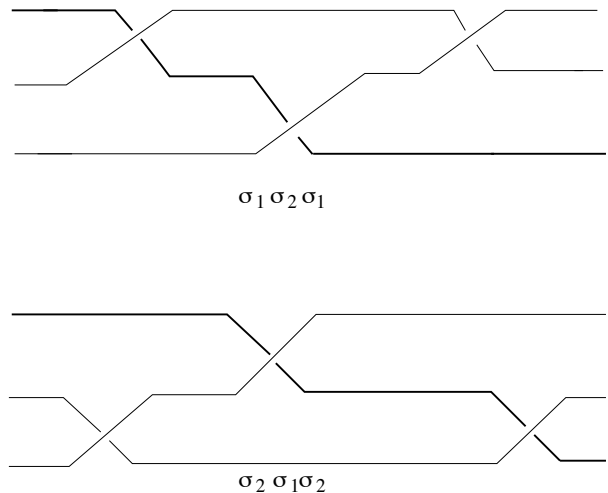}
\caption{The elementary acts of crossing one particle trajectory over another generate the braid group.    The Yang-Baxter relation $\sigma_1 \sigma_{2} \sigma_1 = \sigma_{2}\sigma_1 \sigma_{2}$, made visible here, is its characteristic constraint.}
\end{figure}

If we add to the relations that define the braid group the additional relations
\begin{equation}\label{squareTriviality}
\sigma_j^2 \, = \, 1
\end{equation}
then we arrive at the symmetric (permutation) group $S_N$.   In 3 spatial dimensions, there are more 
ways to untangle trajectories.   Indeed, one can always untangle two world-lines by escaping into the transverse direction to avoid potential intersections, so the permutation of endpoints captures all the topology.    

Yet in 3 dimensions, famously, rotations through $2\pi$ are not topologically trivial.   This topological fact underlies the possibility of spin-$\frac{1}{2}$ (projective) representations of the rotation group.   In such representations, the action of a $2\pi$ rotation is to multiply the wave function by $-1$.  On the other hand,  rotations through $4\pi$ are topologically trivial.   This suggests that for particles with extended structure, that cannot be adequately represented as simple points (e.g., magnetic monopoles, or solitons with extended zero-modes) we should consider relaxing Eqn. (\ref{squareTriviality}) to  
\begin{equation}\label{quadTriviality}
\sigma_j^4 \, = \, 1
\end{equation}
since $\sigma_j^2$ can be implemented by a $2\pi$ rotation moving the particles $(j, j+1)$ around one another, and $\sigma_j^4$ by a $4\pi$ rotation.  
The relations Eqn. (\ref{quadTriviality}), together with Eqns. (\ref{distantCommutation}, \ref{YangBaxter}), define a group intermediate between the braid group and the symmetric group.   

\section{Abelian Anyons}

The substitution 
\begin{equation}\label{anyonRepresentation}
\sigma_j \, \rightarrow \, e^{i\theta}
\end{equation}
preserves the defining relations of the braid group for any phase factor $e^{i \theta}$, so it generates a unitary representation of the braid group.   Thus, at the level of quantum kinematics, it is consistent to weight the amplitudes from topologically distinct classes of trajectories with the corresponding phase factors.   (Of course, the additional constraint Eqn. (\ref{squareTriviality}) reduces the freedom to $e^{i \theta} = \pm 1$.)  This possibility defines the classic, abelian anyons.

There is a simple dynamical realization of anyons, using flux and charge.   Consider a $U(1)$ gauge theory that has particles of charge $q$, associated with a field $\eta$ and is spontaneously broken by a condensate associated with a field of $\rho$ of  charge $mq$, with $m$ an integer\footnote{If $m$ is irrational the gauge group is not compact, i.e. it is the additive group ${\bf R}^+$ rather than $U(1)$.} Gauge transformations that multiply $\eta$ by $e^{2\pi ik/m}$ will multiply $\rho$ by $e^{2\pi i k}$.  Thus for integer $k$ they will leave the condensate invariant, but generally act nontrivially on $\eta$.   We are left with an unbroken gauge group $Z_m$, the integers modulo $m$.   No conventional long-range gauge interaction survives the symmetry breaking, but there is a topological interaction, as follows:
  
The theory supports vortices with flux quantized in units of
\begin{equation}
\Phi_0 \, = \, \frac{2\pi}{mq}
\end{equation}
in units with $\hbar \equiv 1$.   
A particle or group of particles with charge $bq$ moving around a flux $\Phi$ will acquire a phase
\begin{equation}
\exp ibq(\oint dt \vec v \cdot \vec A)  \, = \,  \exp ibq(\oint d\vec x  \cdot \vec A ) \, = \, e^{i \Phi bq} 
\end{equation}
If the flux is $a\Phi_0$, then the phase will be $e^{2\pi i \frac{ab}{m}}$.   

Composites with $(\rm {flux}, \rm {charge}) = (a \Phi_0, bq)$ will be generally be anyons: as we implement the interchange $\sigma_j$, each charge cluster feels the influence of the others flux.  (Note that in two dimensions the familiar flux tubes of three-dimensional physics degenerate to points, so it is proper to regard them as particles.)    There are also topological interactions, involving similar accumulations of phase, for non-identical particles.   What matters are the quantum numbers, or more formally the superselection sector, not the detailed structure of the particles or excitations involved.   

The phase factors that accompany winding have observable consequences.   They lead to a characteristic ``long range'' contribution to the scattering cross-section\footnote{It diverges at small momentum transfer and in the forward direction.}, first computed by Aharonov and B\"ohm \cite{aharonovB} in their classic paper on the significance of the vector potential in quantum mechanics.    Unfortunately, that cross-section may not be easy to access experimentally for anyons that occur as excitations in exotic states of condensed matter.   

\begin{figure}[ht]\label{figure2}
\includegraphics[width=10cm]{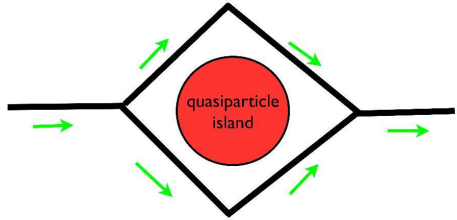}
\caption{A schematic interference experiment to reveal quantum statistics.   We study how the combined current depends on the occupation of the quasiparticle island.}
\end{figure}
Interferometry appears more practical.    The basic concept is simple and familiar, both from optics and (for instance) from SQUID magnetometers.  One divides a coherent flow into two streams, which follow different paths before recombining.   The relative phase between the paths determines the form of the interference, which can range from constructive to destructive recombination of the currents.   We can vary the superselection sector of the area bounded by the paths, and look for corresponding, characteristic changes in the interference.  (See Figure 2.)   Though there are many additional refinements, this is the basic concept behind both the Goldman experiments and other planned anyon detection experiments \cite{plannedExpts}.

Elementary excitations in the fractional quantum Hall effect are predicted to be anyons.   By far the simplest states to analyze are the original Laughlin $1/m$ states, where the excitations are anyons with $\theta = \pi /m $.   There is a rich theory covering more general cases.

\section{Nonabelian Anyons}

The preceding field-theoretic setting for abelian anyons immediately invites nonabelian generalization.   We can have a nonabelian gauge theory broken down to a discrete nonabelian subgroup; vortex-charge composites will then exhibit long range, topological interactions of the same kind as we found in the abelian case, for the same reason.   

The mathematics and physics of the nonabelian case is considerably more complicated than the abelian case, and includes several qualitatively new effects.   First, and most profoundly, we will find ourselves dealing with irreducible {\it multidimensional\/} representations of the braiding operations.   Thus by winding well-separated particles\footnote{From here on I will refer to the excitations simply as particles, though they may be complex collective excitations in terms of the underlying electrons, or other degrees of freedom.}  around one another, in principle arbitrarily slowly, we can not only acquire phase, but even navigate around a multidimensional Hilbert space.   For states involving several  particles, the size of the Hilbert spaces can get quite large: roughly speaking, they grow exponentially in the number of particles.   

As will appear, the states in question are related by locally trivial but globally non-trivial gauge transformations.   Thus they should be very nearly degenerate.   This situation is reminiscent of what one would have if the particles had an internal of freedom -- a spin, say.   However the degrees of freedom here are not localized on the particles, but more subtle and globally distributed.    

The prospect of having very large Hilbert spaces that we can navigate in a controlled way using topologically defined (and thus forgiving!), gentle operations in physical space, and whose states differ in global properties not easily obscured by local perturbations, has inspired visions of {\it topological quantum computing}.   (Preskill \cite{preskill} has written an excellent introductory review.)  The journey from this vision to the level of engineering practice will be challenging, to say the least, but thankfully there are fascinating prospects along the way.   

\begin{figure}[ht]\label{figure3}
\includegraphics[width=11cm]{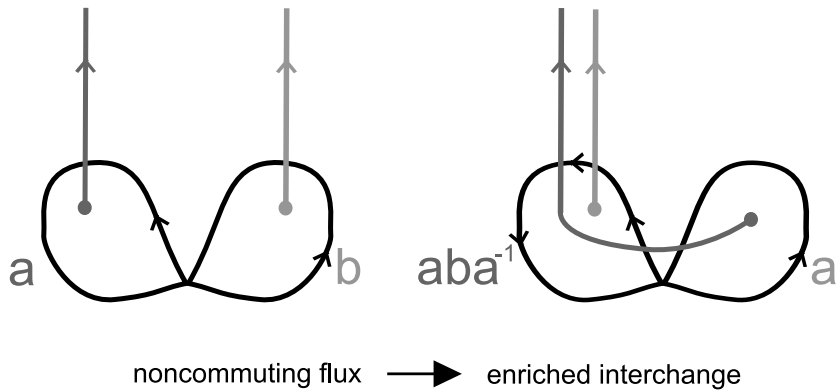}
\caption{By a gauge transformation, the vector potential emanating from a flux point can be bundled into a singular line.   This aids in visualizing the effects of particle interchanges.   Here we see how nonabelian fluxes, as measured by their action on standardized particle trajectories, are modified by particle interchange.}
\end{figure}

The tiny seed from which all this complexity grows is the phenomenon displayed in Figure 3.  To keep track of the topological interactions, it is sufficient to know the total (ordered) line integral of the vector potential around simple circuits issuing from a fixed base point.   This will tell us the group element $a$ that will be applied to a charged particle as it traverses that loop.   (The value of $a$ generally depends on the base point and on the topology of how the loop winds around the regions where flux is concentrated, but not on other details.   More formally, it gives a representation of the fundamental group of the plane with punctures.)   If a charge that belongs to the representation $R$ traverses the loop, it will be transformed according to $R(a)$.   With these understandings, what Figure 3 makes clear is that when two flux points with flux $(a, b)$ get interchanged by winding the second over the first, the new configuration is characterized as $(aba^{-1}, a)$.    Note here that we cannot simply pull the ``Dirac strings'' where flux is taken off through one another, since nonabelian gauge fields self-interact!    So motion of flux tubes in physical space generates non-trivial motion in group space, and thus in the Hilbert space of states with group-theoretic labels.   

\begin{figure}[ht]\label{figure4}
\includegraphics[width=10cm]{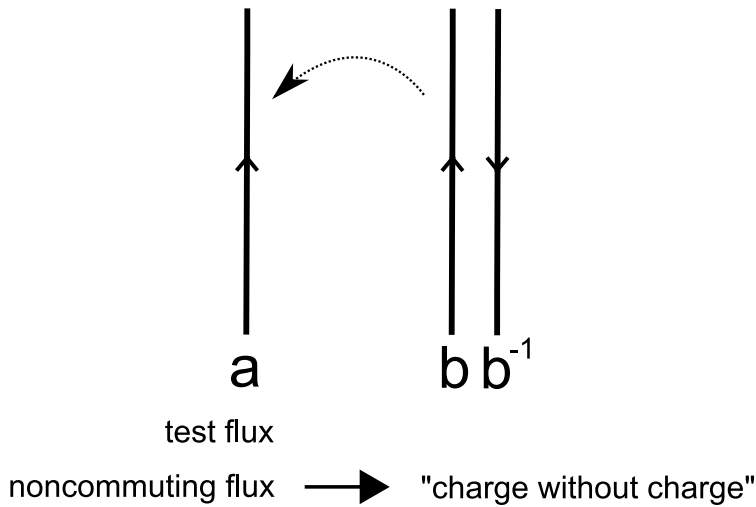}
\caption{Winding a flux-antiflux pair around a test flux, and seeing that it gets conjugated, we learn that the pair carries charge.}
\end{figure}	

As a small taste of the interesting things that occur, consider the slightly more complicated situation displayed in Figure 4, with a pair of fluxes $(b, b^{-1})$ on the right.   It's a fun exercise to apply the rule for looping repeatedly, to find out what happens when we take this pair all the way around $a$ on the right.   One finds
\begin{equation}
(a, (b, b^{-1})) \,  \rightarrow \, (a, (aba^{-1}, a b^{-1} a^{-1}))
\end{equation}
i.e., the pair generally has turned into a different (conjugated) pair.   Iterating, we eventually close on a finite-dimensional space of different kinds of pairs.    There is a non-trival transformation ${\tilde R}(a)$ in this space that implements the effect of the flux  $a$ on pairs that wind around it.  But this property -- to be transformed by the group operation -- is  the defining property of charge!   We conclude that flux pairs -- flux and inverse flux --  act as charges.  We have constructed, as John Wheeler might have said, Charge Without Charge.   

This flux construction makes it clear that nonabelian statistics is consistent with all the general principles of quantum field theory.   Physical realization in condensed matter is a different issue -- in that context, nonabelian gauge fields don't come readily to hand.   Fortunately, and remarkably, there may be other ways to get there.  At least one state of the quantum Hall effect, the so-called Moore-Read state at filling fraction $\frac{5}{2}$, has been identified as a likely candidate to support excitations with nonabelian statistics.   

The nonabelian statistics of the Moore-Read state is closely tied up with spinors \cite{nayakFW} \cite{ivanov}.   I'll give a proper discussion of this, including an extension to 3 + 1 dimensions, elsewhere \cite{halflings}.   Here, I'll just skip to the chase.   Taking $N$ $\gamma_j$ matrices satisfying the usual Clifford algebra relations
\begin{equation}
\{ \gamma_j , \gamma_k \} \, = \, 2\delta_{jk}
\end{equation}
the braiding $\sigma_j$ are realized as
\begin{equation}
\sigma_j \, = \, e^{i\pi / 4} \frac{1}{\sqrt 2} (1 + \gamma_j \gamma_{j+1})
\end{equation}
It's an easy exercise to show that these obey Eqns. (\ref{distantCommutation}, \ref{YangBaxter}), and 
$\sigma_j^4 =1$ (Eqn. (\ref{quadTriviality})) but not $\sigma_j^2 =1$ (Eqn. (\ref{squareTriviality})).

\end{document}